\documentclass[10pt,journal,compsoc]{IEEEtran}
\usepackage{booktabs} 
\usepackage{graphicx}
\usepackage{xspace}
\usepackage[normalem]{ulem}
\useunder{\uline}{\ul}{}
\usepackage{multirow}
\usepackage{lscape}
\usepackage{rotating}
\usepackage{supertabular}
\usepackage{longtable}
\usepackage{enumerate}
\usepackage{comment}
\usepackage{ragged2e}
\usepackage{array}
\PassOptionsToPackage{hyphens}{url}
\usepackage{url}
\usepackage[breaklinks]{hyperref}
\usepackage{array}
\usepackage{ragged2e}
\usepackage{caption}
\newcolumntype{P}[1]{>{\RaggedRight\hspace{0pt}}p{#1}}
\usepackage{fontawesome}
\usepackage[table,xcdraw]{xcolor}
\usepackage{tcolorbox}
\usepackage[switch]{lineno}

\newif\ifdraft
\drafttrue

\ifCLASSINFOpdf
\else
\fi
\begin{document}
\bstctlcite{IEEEexample:BSTcontrol}
%
\title{Human Values in Mobile App Development: An Empirical Study on Bangladeshi Agriculture Mobile Apps}

\author{Rifat~Ara~Shams,~\IEEEmembership{}
        Mojtaba~Shahin,~\IEEEmembership{}
        Gillian~Oliver,~\IEEEmembership{}
        Jon~Whittle,~\IEEEmembership{}
        Waqar~Hussain,~\IEEEmembership{}\\
        Harsha~Perera,~\IEEEmembership{}
        and~Arif~Nurwidyantoro~\IEEEmembership{}
\IEEEcompsocitemizethanks{\IEEEcompsocthanksitem 
Rifat Ara Shams, Mojtaba Shahin, Gillian Oliver, Waqar Hussain, Harsha Perera and Arif Nurwidyantoro are with the Faculty of Information Technology, Monash University, Australia.
\protect\\
E-mail: \{Rifat.Shams,~Mojtaba.Shahin,~Gillian.Oliver,~Waqar.Hussain,~Harsha.Perera,~Arif.Nurwidyantoro\}@monash.edu}\\
\IEEEcompsocitemizethanks{\IEEEcompsocthanksitem 
Jon Whittle is with CSIRO's Data61, Australia.
\protect\\
E-mail: Jon.Whittle@data61.csiro.au
}

\thanks{Manuscript submitted to IEEE Transactions on Software Engineering (2021)}
}


%


\IEEEtitleabstractindextext{
\begin{abstract}
\justifying
Given the ubiquity of mobile applications (apps) in daily lives, understanding and reflecting end-users' human values (e.g., \textit{transparency}, \textit{privacy}, \textit{social recognition} etc.) in apps has become increasingly important. Violations of end users' values  by software applications have been reported in the media and have resulted in a wide range of difficulties for end users. Value violations may bring more and lasting problems for marginalized and vulnerable groups of end-users. This research aims to understand the extent to which the values of Bangladeshi female farmers, marginalized and vulnerable end-users, who are less studied by the software engineering community, are reflected in agriculture apps in Bangladesh. Further to this, we aim to identify possible strategies to embed their values in those apps. To this end, we conducted a mixed-methods empirical study consisting of 13 interviews with app practitioners and four focus groups with 20 Bangladeshi female farmers. The accumulated results from the interviews and focus groups identified 22 values of Bangladeshi female farmers, which the participants expect to be reflected in the agriculture apps. Among these 22 values, 15 values (e.g., \textit{accuracy}, \textit{independence}) are already reflected and 7 values (e.g., \textit{accessibility}, \textit{pleasure}) are ignored/violated in the existing agriculture apps. We also identified 14 strategies (e.g., \textit{``applying human-centered approaches to elicit values"}, \textit{``establishing a dedicated team/person for values concerns"}) to address Bangladeshi female farmers' values in agriculture apps.

\end{abstract}

\justifying
}

\maketitle
\IEEEpeerreviewmaketitle
\section{Introduction}
\label{sec:introduction}

Software is an integral part of human lives and thus it is essential to consider human values such as \textit{security}, \textit{privacy}, \textit{social justice}, \textit{freedom}, \textit{independence}, \textit{fairness}, \textit{accessibility}, and \textit{tradition} in software design. Human values are defined as "desirable, trans-situational goals, varying in importance, that serve as guiding principles in people’s lives" \cite{schwartz2013value}. Software practitioners often design software based on their knowledge of desired functional and non-functional requirements and ignore human values, resulting in failure to meet users' expectations \cite{curumsing2019emotion}. As a result, the developed software causes users' disappointment and negative socio-economic consequences \cite{mougouei2018operationalizing}. However, once human value violations occur in software systems, the importance of considering human values is more vividly demonstrated and publicised. For example, a self-harming game named ``Blue Whale Challenge" was responsible for the death of 153 teenagers around the world \cite{gowda2019blue}. In this game, the teenagers were given 50 deadly tasks to complete in 50 days with committing suicide as the final task \cite{mukhra2019blue}. Therefore, this game arguably violated society's values such as \textit{preservation of life}. 
Another value violation was taken place by the facial-recognition software of Google photos. Google photos tagged a photo of two black people under the label, ``gorillas" which was extremely offensive and violated the people's values such as \textit{recognition} and \textit{public image} \cite{gorilla}. 
In another example, a recent study showed that almost 61\% (22 out of 36) menstruation apps shared users' sensitive personal information with Facebook immediately after opening the app without the consents of the users \cite{Nobody2019, al2020we}. This kind of \textit{privacy} breach can have a negative impact on women's  mental health.

Despite the significance of recognising human values in software, current Software Engineering (SE) research and practice do not sufficiently consider human values \cite{mougouei2018operationalizing}. Only 16\% of 1350 papers published between 2015 and 2018 in top-tier SE journals and conferences were directly relevant to human values \cite{perera2020study}. However, some  recent research in software engineering has focused on human values. For example, Mougouei et al. explained the importance of human values in software engineering \cite{mougouei2018operationalizing}. Shams et al. looked at app reviews to identify the present and ignored/violated values in apps \cite{shams2020society}. Obie et al. developed an approach to detect values violations reported in app reviews \cite{obie2021first}. Furthermore, some techniques have been proposed to integrate human values in software-intensive systems development, such as Value-Based Requirements Engineering (VBRE) \cite{thew2018value}, Value-Sensitive Design (VSD) \cite{davis2015value}, Value-Sensitive Software Development (VSSD) \cite{aldewereld2015design}, Values-First SE \cite{ferrario2016values}, and Values Q-sort \cite{winter2018measuring}.  

In \cite{hussain2020human}, we conducted  two case studies of software companies to identify the perceptions of human values among practitioners, to explore how the two companies address human values in software engineering practice, and to identify the challenges they face in addressing human values in software \cite{hussain2020human}. In another study, we conducted a case study on how and where human values can be addressed in the Scaled Agile Framework (SAFe), leading to identifying five intervention points (artefacts, roles, ceremonies, practices, and culture) \cite{hussain2021can}.

The studies mentioned above have not collected data about the perceptions of end-users on human values, which is an essential factor to design values-based software (software that reflects end-users' values) \cite{kujala2009value}. In addition, there is little research on human values in SE from the perspective of vulnerable and underrepresented end-users (e.g., women in developing countries) \cite{shams2021measuring}. However, there are a few works in SE and HCI on addressing vulnerable end-users' needs \cite{curumsing2019emotion, mutanu2020integrating, elias2018design, vines2013designing, baez2018agile}. 
Finally, there is little if anything known about how to address human values in mobile apps. Mobile apps are rich and complex \cite{stuurman2014design} and developing such apps is different from desktop or enterprise software \cite{longoria2001designing}. For example, the functional requirements, information architecture, and layout of mobile apps are different than other software. Mobile apps also cause more usability and utility problems \cite{longoria2001designing}. Therefore, the challenges faced by app developers and the strategies adopted by them to address values in apps might be different from those of other types of software systems.

This research aims to identify the extent to which mobile apps reflect their end-users' values and explore the possible strategies to address end-users' values in mobile apps. We conducted a mixed-methods empirical study, which collected data from four focus groups with end-users of Bangladeshi agriculture mobile apps and 13 interviews with practitioners involved in developing those mobile apps. More specifically, 20 Bangladeshi female farmers, who regularly use agriculture mobile apps in their daily agricultural activities, attended the focus groups. In this paper, we used the terms ``Bangladeshi female farmers" and ``farmers" interchangeably. Our findings indicate:

\begin{itemize}
\item 22 values of Bangladeshi female farmers, which the focus groups' participants and interviewees expect to be reflected in the agriculture apps. Among them, 15 values (e.g., \textit{accuracy}, \textit{independence}) are reflected (present) and 7 values (e.g., \textit{accessibility}, \textit{pleasure}) are ignored/violated (missing) in the existing agriculture apps. 
\item 14 strategies (e.g., \textit{``applying human-centered approaches to elicit values"}, \textit{``establishing a dedicated team/person for values concerns"}) to address Bangladeshi female farmers' values in agriculture apps.
\end{itemize}

\textbf{Paper Organization:} Section \ref{sec:background} discusses the background of this research and the related work. In Section \ref{sec:methodology}, we briefly explain our research method. Section \ref{sec:results} reports the results of this study which we discuss in Section \ref{sec:discussion_v2}. Section \ref{sec:ttv} discusses the possible threats to validity of this research. Finally, we conclude our research with possible future research directions in Section \ref{sec:conclusion}.

\section{Background and Related Work}
\label{sec:background}

\subsection{Human Values and Values Theory}
Human values are defined as ``enduring beliefs that a specific mode of conduct or end state of existence is personally or socially preferable to an opposite or converse mode of conduct or end state of existence" \cite{rokeach1973nature}. Therefore, human values refer to people's personal and social preferences \cite{braithwaite1985structure} that reflect on attitudes and behaviors of individuals, and the functioning of organizations, institutions, and societies \cite{schwartz2001extending}. Values have been used 
to identify personality and to understand what is and is not important in life \cite{allport1960study}. According to the social scientist, Shalom H. Schwartz, ``values are beliefs linked to emotions, refer to desirable goals that motivate action, transcend specific actions and situations, serve as standards for evaluating actions, policies, people, and events, and form a relatively enduring hierarchical system ordered by importance" \cite{schwartz1992universals}.

\begin{figure*}[!htbp]
        \centering
        \includegraphics[width=0.8\textwidth]{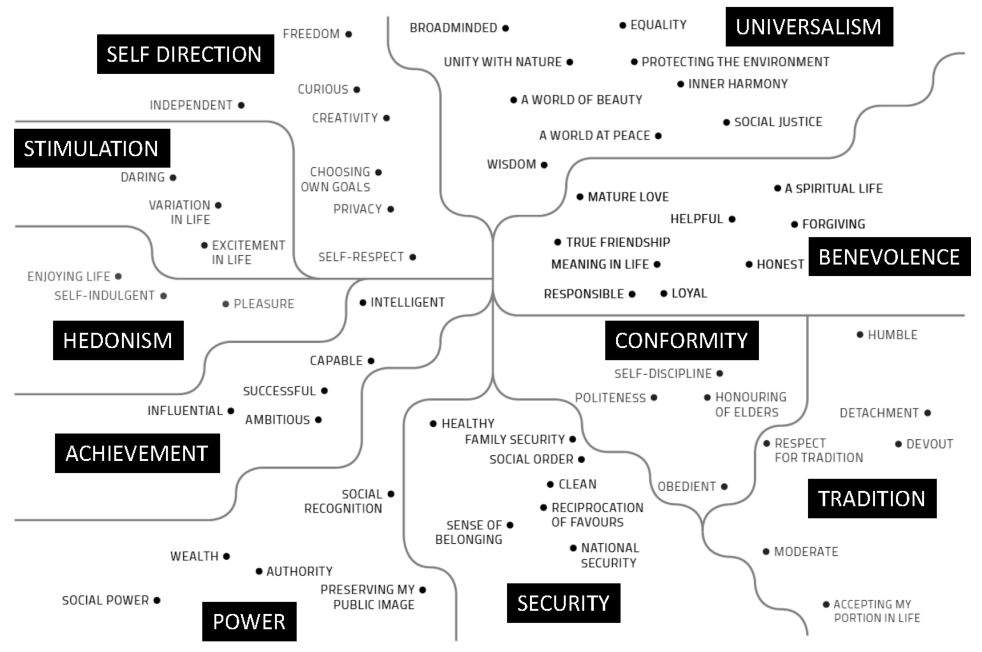}
        \caption{Schwartz' theoretical model of basic human values \cite{schwartz1992universals}}
        \label{fig:SchwartzTheory}
    \end{figure*}

Social scientists have been conducting research to conceptualize human values since 1931 \cite{inbook}. The first theory of values was proposed by Allport and Vernon in 1931 based on personality typology \cite{vernon1931test}. After decades, in 1960, Allport et al. classified values into six categories: theoretical, economic, aesthetic, social, political, and religious \cite{allport1960study}. In 1973, Rokeach proposed a value model where he categorized values into goals in life and mode of conduct which he named terminal values and instrumental values respectively \cite{rokeach1973nature}. In 1980, Hofstede divided values into two categories, desired and desirable i.e. what people actually desire and what they think ought to be desired respectively \cite{hofstede1980culture}. In 1992, Schwartz introduced his theory of basic human values which recognized 10 human values defined by their motivational goals and measured from 58 value items (see \autoref{fig:SchwartzTheory}) \cite{schwartz1992universals}. In 2004, Parashar et al. divided values into two concepts: individual behavior and cultural practices which they named micro and macro values \cite{parashar2004perception}. In 2008, Inglehart identified two dimensions of values according to the cross-cultural variations:  post-materialist and self-expression values \cite{inglehart2008changing}. In 2010, Cheng and Fleischmann developed a meta-inventory of human values by analyzing 12 value inventories \cite{cheng2010developing}. In 2014, Gouveia et al. proposed a framework with six basic value categories (existence values, promotion values, normative values, suprapersonal values, excitement values, interactive values) and three specific value items under each category \cite{gouveia2014functional}.

In this research, we used Schwartz theory of basic human values while exploring Bangladeshi female farmers' values, as it is the most cited and widely used values theory \cite{shams2020society} in social science along with other areas such as computer science \cite{barcelo2014social} and software engineering \cite{ferrario2014software}. However, we did not limit ourselves to Schwartz's values, and we kept ourselves open to explore other values which were not identified in Schwartz's theory.

\subsection{Values in Technology Design}
Human values have been of interest in technology design and development since the 1970s \cite{van2015design}. Batya Friedman was the first to propose an approach to embed values in technology design, namely Value Sensitive Design (VSD) \cite{van2015design}. According to Friedman et al., ``Value Sensitive Design is a theoretically grounded approach to the design of technology that accounts for human values in a principled and comprehensive manner throughout the design process." \cite{friedman2008value}. In \cite{friedman2007human}, Friedman et al. reviewed and summarized different approaches, projects, and ideas to embed human values and ethics in technology design practice. Friedman et al. also explained conceptual, empirical, and technical investigations of VSD and also provided suggestions to use VSD \cite{friedman2008value}.

Other than VSD, there are a few related approaches to embed values in technology design. For example, value conscious design \cite{manders2011values}, values in design \cite{flanagan2005values}, and design for values \cite{van2015design}. According to Huits, "VSD lacks complimentary or explicit ethical theory for dealing with value trade-offs" and therefore, she proposed to complement VSD with an explicit and justified ethical principle which she named Value Conscious Design \cite{manders2011values}. Another approach, Values in Design, commits to design systems that addresses values of the designers, users, other stakeholders, and the surrounding society \cite{flanagan2005values}. Values in Design understands values and technology in the early stages of design with practical definitions of the terms, “values” and “design” by practical exercises \cite{knobel2011values} such as Values at Play \cite{flanagan2005values} and Envisioning Cards \cite{friedman2012envisioning}. On the other hand, Design for Values consists of three features: values are imparted to technology, conscious and explicit thinking about the values is morally significant, and values need to be considered at the early stage of design and development of technology \cite{van2015design}.
    
\subsection{Values in Software Engineering}
    
    \subsubsection{\textbf{Values in Software Development Process}}
    Human values in software engineering is an emerging area. Mougouei et al. developed a research roadmap for human values in SE where the research gaps of tracing and measuring values in SE were identified \cite{mougouei2018operationalizing}. Ferrario et al. introduced a Values-First SE framework to consider intrinsic and extrinsic values during SE decision making processes \cite{ferrario2016values}. Winter et al. designed and developed Values Q-Sort, a value measurement tool, to investigate values at system, personal, and instantiation levels of SE \cite{winter2018measuring}.
    
    A few studies have taken human values into account during different phases of software development life cycle (SDLC). For example, Perera et al. introduced a framework named Continual Value(s) Assessment (CVA) to integrate, trace, and evaluate human values throughout the SDLC \cite{perera2020continual}. Thew and Sutcliffe proposed a taxonomy and analysis method named Values Based Requirement Engineering (VBRE) that complements existing analysis of non-functional requirements by eliciting stakeholders' values, motivations, and emotions \cite{thew2018value}. Hussain et al. proposed a framework to specify the value implications of software design patterns as well as to guide on the value-conscious adoption of design patterns \cite{hussain2018integrating}. Aldewereld et al. introduced a conceptual framework, Value-Sensitive Software Development (VSSD) as a design for values approach to the development of software that translated the abstract values into more concrete elements \cite{aldewereld2015design}.
    
    \subsubsection{\textbf{Values in Agile Methods}}
    In Agile development methods, software engineers should make a decision that takes into account both technical details and human values \cite{miller2005agile}. Miller et al. defined computer ethics as technical decisions considering human values. For this purpose, they recommended two ethical analysis techniques, utilitarian ideas and deontological ideas, to know the advantages and disadvantages of Agile software development in terms of computer ethics \cite{miller2005agile}. Hussain et al. revealed five intervention points (artefacts, roles, ceremonies, practices, and culture) in Scaled Agile Framework (SAFe) where human values should be considered \cite{hussain2021can}. Schön et al. identified four methodologies that make Agile software development more human-centric: Human-Centered Design, Design Thinking, Contextual Inquiry, and Participatory Design \cite{schon2017agile}.
    
    \subsubsection{\textbf{Values in Apps}}
    Recently, a few research studies have been conducted on values in mobile applications. For example, Value-Sensitive Design (VSD) has been applied in the analysis of user reviews of diabetes apps to identify patients' values and propose design techniques to consider their values \cite{dadgar2015diabetes}. In \cite{shams2020society}, Shams et al. manually analyzed user reviews of Bangladeshi agriculture apps to explore the desired, present and ignored/violated values of the users. Obie et al. used natural language processing techniques to analyze app reviews to detect perceived end user human values violations \cite{obie2021first}. Rowe raised the issue of values dilemmas in COVID-19 contact tracing apps and proposed possible ways to handle this dilemma \cite{rowe2020contact}. 
    
\subsection{Bangladesh, Women in Agriculture and Smartphone}
Bangladesh is a country in South-East Asia with rich fertile land and many rivers due to its geographical location \cite{chowhan2020role}. It made Bangladesh an agriculture-dependent country where agriculture contributes 14.23\% of GDP in 2019 \cite{chowhan2020role}. 
In Bangladesh, more than 80\% of the total population depends on agriculture to earn their livelihood (2016) \cite{faroque2021effect} and
38\% labor force works in the agricultural sector (2019) \cite{WorldBankEmployment2021}. However, women's participation in agricultural activities has increased over time, resulting in the ``\textit{feminization of agriculture}" \cite{jaim2011women}. 
With the urbanization process, a significant number of rural labor force migrates to the urban areas, resulting in women taking over the traditional agricultural activities of men \cite{de2021migration}. In Bangladesh, women's participation in agriculture has started expanding from 1999/2000, less than 20\% of the agricultural labour force in 1999/2000, 33.6\% in 2010 \cite{sraboni2014empowered}, more than 50\% in 2016 \cite{FAO2016}, and 58\% in 2019 \cite{WorldBankFemaleAgri2021}.

Mobile phones play an important role in agriculture as they provide the  opportunity for farmers to access information regarding agricultural initiatives \cite{stillman2020after}. To date, there are approximately 3.2 billion smartphones users across the world \cite{blair2021} and 92.8\% of the mobile phone users use Internet via smartphones \cite{DataReportal2021}. With the increase of the use of Internet, mobile apps have become increasingly  popular. There are 4.66 million apps available both in the Google Play Store (2.87 million) and the Apple App Store (1.96 million) \cite{blair2021}. Nowadays, agriculture apps are gaining popularity as they play a vital role in agricultural development. In 2018, there were 35 Bangladeshi agriculture apps available in Google play \cite{shams2020society} used by all the stakeholders involved in agriculture including farmers \cite{rahman2020utility, shams2021measuring}. These apps provide effective means of getting feedback and communication, such as sharing knowledge, information, and field-related problems \cite{chowhan2020role}. Bangladeshi farmers use these apps for weather forecasting and to obtain agricultural information \cite{chowhan2020role} such as agricultural problem solving, detecting diseases, finding disease solutions, and using the recommended medicines \cite{roshidul2016potential, kundu2017smart}.

There are a few studies on the Bangladeshi agriculture apps, such as the role of mobile apps on Bangladesh agriculture and its future scope \cite{chowhan2020role}, farmers' empowerment through agriculture mobile apps \cite{sharma2019ikhedut}, and exploring the techniques suitable to develop agriculture apps in local languages to identify diseases and management of maize crop in Bangladesh \cite{roshidul2016potential}. However, there is limited research on Bangladeshi farmers' values in agriculture apps. In \cite{shams2021measuring}, Shams et al. identified the values of Bangladeshi female farmers. In another study, \cite{shams2020society}, Shams et al. explored the present and missing values of the users of Bangladeshi agriculture apps by analyzing user reviews. In this research, we determine which of the values of Bangladeshi female farmers are present and which ones are missing in existing agriculture apps. We also identify strategies to address their values in apps from both the end-users' and app practitioners' perspectives.

\section{Methodology}
\label{sec:methodology}

With the goal to address Bangladeshi female farmers' values in agriculture mobile apps, we formulated the following two research questions.

   \noindent \textit{\textbf{RQ1.} To what extent do existing agriculture mobile apps reflect Bangladeshi female farmers' values?}
   
   \noindent \textit{\textbf{RQ2.} What are the strategies to address Bangladeshi female farmers' values in agriculture mobile apps?} 
    
\begin{figure*}[!htbp]
            \centering
            \includegraphics[width=0.9\textwidth]{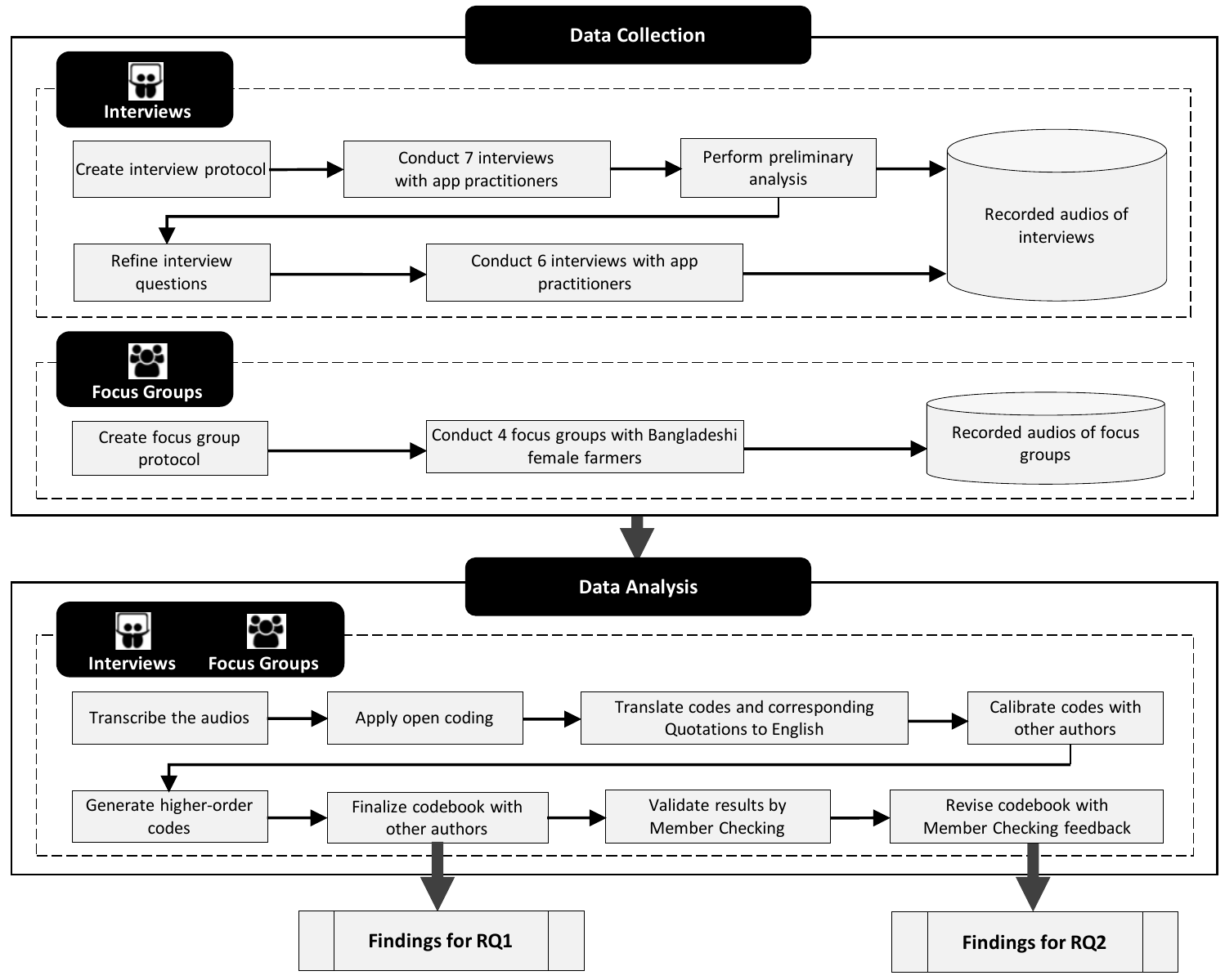}
            \caption{An overview of the research method}
            \label{fig:method}
    \end{figure*}

We conducted a mixed-methods empirical study, which collected data through 13 semi-structured interviews with Bangladeshi agriculture mobile app practitioners and 4 focus groups with 20 Bangladeshi female farmers to answer the research questions. \autoref{fig:method} shows an overview of our research method. As this study worked with humans, ethics approval with project ID 21891 was acquired from Monash University Human Research Ethics Committee (MUHREC) on 21/11/2019.

\subsection{Interviews}

\subsubsection{\textbf{Protocol}}
We conducted 13 semi-structured interviews \cite{kallio2016systematic} with agriculture mobile app practitioners. The first author conducted all the interviews. The interviews consisted of three parts. In the first part, 5-10 minutes were allocated for ice-breaking. In this part, the first author introduced herself and conducted an informal conversation to put the interviewees at ease. In the second part, the interviewees were informed of the objective of the research project and the structure of the interview. In addition, the interviewer explained the possible benefits and risks of this research and the actions taken to ensure data confidentiality and the safe storage and use of collected data. The interviewees were provided with an explanatory statement and asked to sign a consent form for this purpose. The last part took 27-66 minutes where the participants were asked 21 questions that could be classified into three groups:
\begin{itemize}
    \item \textbf{Demographic questions}: We asked 3 demographic questions (i.e., What is your current role and the associated responsibilities? How long have you been working in this organization? How long have you been working in the software industry?).
    \item \textbf{Reflecting values in apps}: We asked 6 questions to understand the extent to which Bangladeshi female farmers' values are reflected in the existing apps (e.g., Which aspects/features of the existing mobile apps are (more) important for the female farmers in their personal lives or make them, for example, happy, powerful, independent, etc. in their personal lives? Why?).
    \item \textbf{Strategies to address values in apps}: Through 12 questions (e.g., How software development process needs to get updated to meet the current demands for making impactful apps?), interviewees were asked to share any strategies (e.g., practices, tools, or techniques) that they think helpful or have used to address human values in agriculture apps.
\end{itemize}

Given that both the interviewer and interviewees were Bangladeshi, all interviews were conducted in their mother tongue (Bengali) to facilitate communication and make both sides comfortable. The interviews were conducted in two rounds with different participants for each round. In the first round, seven interviews were carried out at the interviewees' workplaces in Bangladesh from December 2019 to February 2020. Before proceeding to the second round of the interviews, we undertook preliminary analysis of the data collected from the first round. This helped us refine the wording of some interview questions (e.g., those that were vague) and remove two unnecessary or redundant questions (see the interview questionnaire \cite{replpack}). In the second round, six interviews were conducted via Zoom due to the COVID-19 pandemic.
As the interviews were semi-structured, we asked many follow-up questions depending on the interviewees' answers. All the 13 interviews were audio-recorded and stored in Google drive for the data analysis.

\subsubsection{\textbf{Participants}}
The main criterion to select the interviewees was finding practitioners currently are or previously were involved in agriculture app development for Bangladeshi female farmers. 
Our research contributes to an ICT4D project named PROTIC, which works for the resilience of Bangladeshi female farmers through the use of ICT \cite{stillman2020after}. PROTIC is a joint project between Oxfam Bangladesh (a multi-national charitable organization) and Monash University. Oxfam Bangladesh has had a long history of involvement with agricultural app developers, so was ideally placed to provide initial contacts. Oxfam introduced 5 practitioners to us after informing them the purpose of our research. We recruited 8 other interviewees by applying Snowball Sampling Method (SSM) \cite{cohen2011field}.

\textbf{Interviewees characteristics}: 7 (53.8\%) males and 6 (46.2\%) females participated in the interviews. Of the 13 interviewees, 2 (15.38\%) had more than 20 years, 1 (7.69\%) had 16-20 years, 3 (23.08\%) had 11-15 years, 4 (30.77\%) had 6-10 years, and 3 (23.08\%) had less than 6 years of experiences in software development/research sectors. The demographics of the interviewees are shown in \autoref{tbl:interviewees}.

\begin{table*}
\centering
\caption{Interviewee demographics (M= Male, F= Female)}
\label{tbl:interviewees}
\resizebox{\textwidth}{!}{%
\renewcommand{\arraystretch}{1.3}
\begin{tabular}{|c|P{4.5cm}|P{9cm}|P{5cm}|c|c|}
\hline
\textbf{ID} & \textbf{Role}                    & \textbf{Responsibilities}                                                                                                                                         & \textbf{Organization Information}    & \textbf{Gender} & \textbf{Year of Exp.} \\\hline
I1          & Software Developer/ Technical Lead                   & Leading teams to develop agriculture apps                                                                                                                                             & Software Company Focused on Agricultural Development     & M               & 24                    \\\hline
I2          & Software Developer                                   & Analysing data, developing weather contents of agriculture apps                                                                                                                       & Software Company Focused on Agricultural Development     & M               & 11                    \\\hline
I3          & Chairman and CEO                                     & Curating contents in agriculture apps, decision making, communicating with advisory panel                                                                                             & Software Company Focused on Agricultural Development     & F               & 15                    \\\hline
I4          & Senior Program Officer                             & Fieldwork, analysing end users' requirements and feedback, communicating with the developers                                                                                           & Multi-national Charitable Organization                   & F               & 9                     \\\hline
I5          & Project Officer in ICT and Development               & Fieldwork, analysing end users' requirements and feedback, communicating with the developers                                                                                           & Multi-national Charitable Organization                   & F               & 4                     \\\hline
I6          & Project Officer in ICT and Development               & Managing ICT intervention, fieldwork, analysing end users' requirements and feedback, communicating with the developers                                                                & Multi-national Charitable Organization                   & F               & 7                     \\\hline
I7          & Senior Program Manager                             & Fieldwork, analysing end users' requirements and feedback, communicating with the developers                                                                                           & Multi-national Charitable Organization                   & M               & 21                    \\\hline
I8          & Software Developer                                   & Analysing requirement, creating app contents, developing agriculture apps                                                                                                              & Agriculture University                                   & F               & 2                     \\\hline
I9          & Software Developer/ Researcher on ICT in Agriculture & Research on agricultural development through ICT, fieldwork, collaborating and communicating with project partners and team members, developing agriculture apps                      & Institute of ICT and Development & F               & 6                     \\\hline
I10         & Software Developer                                   & Analysing requirements, creating app contents, developing agriculture apps                                                                                                             & Agriculture University                                   & M               & 7                     \\\hline
I11         & Software Developer/ Researcher on ICT in Agriculture & Leading agriculture apps development team, researching on agriculture app development                                                                                                 & Agriculture University                                   & M               & 2                     \\\hline
I12         & CEO/ Researcher on ICT in Agriculture                & Researching on agricultural development through ICT, developing apps for farmers and youth, helping women entrepreneur through ICT                            & Institute of ICT and Development & M               & 13                    \\\hline
I13         & Software Developer and CEO                           & Designing UX by using human-centered design, qualitative and quantitative research on apps, working for the marginalized women (particularly for the Bangladeshi women dairy farmers) & Software Company Focused on Human-Centered Design        & M               & 20                   
    \\ \hline
\end{tabular}
}
\end{table*}

\subsection{Focus Groups}

\subsubsection{\textbf{Protocol}}
\begin{figure}[!htbp]
        \centering
        \centering \includegraphics[width=0.48\textwidth]{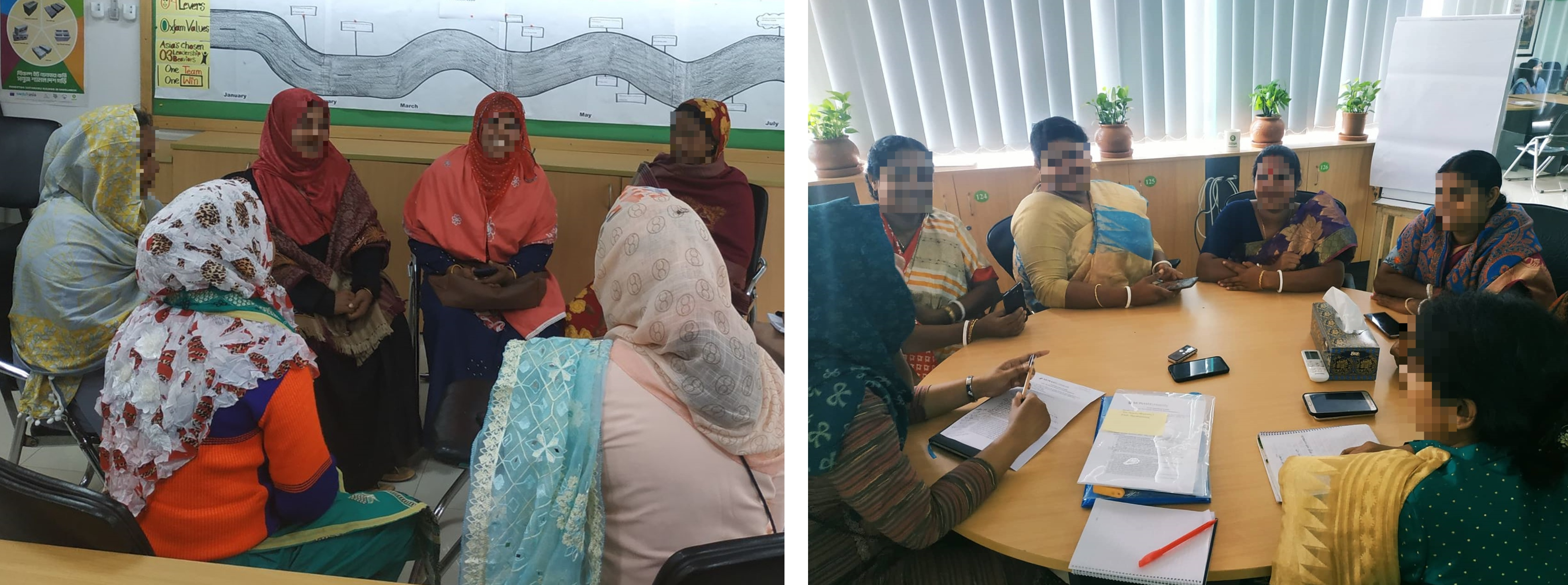}
        \caption{Conducting focus groups}
        \label{fig:focus_groups_conduction}
    \end{figure}

We conducted 4 focus groups with a total of 20 Bangladeshi female farmers who use agriculture apps to support their farming activities. 10 of them were from North region and 10 from South. The first author carried out all 4 focus groups in Bengali. Five female farmers from one region participated in each focus group. All four focus groups were conducted in two days at the Oxfam Bangladesh office. The participants were provided a comfortable environment with round seating arrangements and refreshments.
In addition, all focus group participants had previously visited the Oxfam Bangladesh office several times, so the setting was familiar and helped them feel at ease. \autoref{fig:focus_groups_conduction} shows two pictures of conducting focus groups in two different days.

Each focus group consisted of three parts. The first part functioned as an ice-breaker; the first author met the participants before data collection started and spent one hour with them chatting to help put them at their ease. The second part covered explaining our research, its objectives, expected outcomes, possible benefits and risks of this research, and the strategies adopted to ensure the confidentiality of the collected data. In this part, the participants were also provided with an explanatory statement of this research and asked to sign the consent form. The last part consisted of 19 questions, which took 58 minutes to 80 minutes (see the focus groups questionnaire \cite{replpack}). More specifically, we asked the following:
\begin{itemize}
    \item \textbf{Agriculture activities}: We asked 2 general questions regarding the farmers' agriculture activities (e.g., What kind of farming do you do? How long have you been working as a farmer?). 
    \item \textbf{Values probing:} We asked 7 questions to understand the farmers' values and the importance level of those values for the farmers (e.g., What are the most important things to do in your life? What are the goals to achieve in your life?). 
    \item \textbf{Reflecting values in apps}: 6 questions regarding the reflection of their values in existing agriculture apps were asked (e.g., What kind of help do you usually take from the apps? Can you recall any feature that you think made your life better?).
    \item \textbf{Strategies to address values in apps}: Through 4 questions, we solicited the farmers' opinions on how they think their values can be better addressed in apps (e.g., What particular features do you expect from the apps? What suggestions do you have to make the agriculture apps more usable for you?). 

\end{itemize}
These 4 focus groups were recorded and recordings stored in Google drive for the data analysis.

\subsubsection{\textbf{Participants}}
The primary selection criterion for identifying participants for the focus groups was choosing Bangladeshi female farmers who use smartphones and agriculture apps in their daily agricultural activities. Two local NGOs connected with Oxfam,  Shushilon and Pollisree, trained 200 female farmers  how to use smartphones and different agriculture mobile applications in their daily activities. After discussing our criterion to select the participants with Oxfam, they chose 10 Bangladeshi female farmers from the North region (sandy area \cite{anik2012climate}) and 10 from the South region (coastal area \cite{rakib2019investigation}) among those 200 trained female farmers and introduced them to the first author. The 20 female farmers were invited to the Oxfam Bangladesh office, where we conducted the focus groups.

\subsection{Data Analysis}
\label{da}
\autoref{fig:method} shows the overview of the data analysis for both interviews and focus groups. We analyzed the data following the five steps as given below.

    \textbf{Transcribing data:} In the first step, the audio files of the interviews and focus groups were given to a Bangladeshi transcriber to transcribe them in Bengali. The transcriber had no previous experience with human values and values in mobile apps; therefore, there was no chance of being biased while transcribing. Once the transcripts were ready, the first author randomly compared the transcripts of three interviews and one focus group with the audio files to ensure that the data was transcribed precisely.
    
    \textbf{Coding data:} In this step, the first author listened to the audios to become familiar with the data. Then, she performed the open coding procedure \cite{khandkar2009open} on each source of transcribed data separately. As we had a high volume of data, we decided to use NVivo software\footnote{ \href{https://www.qsrinternational.com/nvivo-qualitative-data-analysis-software/home}{\url{https://www.qsrinternational.com/nvivo-qualitative-data-analysis-software/home}}} to support open coding, which helped us review and edit the codes and the extracted data under a particular code easily. The open coding process started by reading through the transcripts and making notes to identify all aspects of our research questions by the first author. Then, she went through the transcripts again and analyzed the notes to create codes. All the codes were kept under two higher-order codes: interview codes and focus group codes. Applying this process to each data source led to a preliminary codebook with corresponding code descriptions and quotations. At this stage, the coding was driven by only one Bangladeshi analyst (the first author), as the audio files and transcripts were in Bengali. To mitigate the potential threats due to the absence of investigator triangulation, the first author repeated the same process three times to ensure intra-rater reliability \cite{ergai2016assessment}. 
    
    \textbf{Translating codes and corresponding information:} After creating the codebook, the first author translated the codes with their corresponding quotations and relevant information from Bengali to English.
    
    \textbf{Finalizing the codebook:} In this step, the first author discussed the initial codes, their corresponding quotations, and other relevant information with the second author.  The first and second authors had four discussion sessions (i.e., each took 1 hour) to calibrate the codes. During this time, the first author went back to the raw data several times to recheck and refine the codes. This led to the construction of the second version of the codebook. After that, the first author analyzed the similarities and differences among the codes, generated higher-order codes, and precisely named them (we refer to such higher-order codes as ``groups"). Next, the first author discussed the emerged groups, their corresponding codes, and quotations with the third author. The feedback collected from the third author resulted in changing the names of some of the groups and merging some of the codes. Then, the groups, corresponding codes, and illustrative quotations were shared with the second author again. The first and second authors conducted four more iterations to finalize the groups and the codes under each group. Finally, the results (Section \ref{sec:results}) were shared with other authors, and their comments made some tweaks to the groups and their corresponding codes. The process ended when everybody was satisfied with the results. \autoref{table:open_coding_example} shows examples of our coding process with some raw data of interviews and focus groups.
    
    \begin{table*}
\centering
\caption{Examples of open coding process}
\label{table:open_coding_example}
\resizebox{\textwidth}{!}{%
\renewcommand{\arraystretch}{1.3}
\begin{tabular}{p{1.2cm}|P{6cm}|P{2.6cm}|P{1.8cm}|p{1.8cm}|p{1.6cm}}

\hline
\textbf{Data Source} & \textbf{Raw Data}                                                                                                                                                                                                                                                                           & \textbf{Notes}                                                                             & \textbf{Code}                  & \textbf{Refined Code}                                           & \textbf{Group}                          \\ \hline
Interviews           & \textit{``Maybe \underline{too much information}, \underline{complicated design} and \underline{too much colors} can make the apps attractive but \underline{not usable}. My recommendation is to design the apps in a way that it feels \underline{simple to use}."}                                                                                   & Simple contents and design to make app more usable                                         & Making the app simple          & \multirow{2}{2cm}{Customizing apps based on user literacy levels} & \multirow{2}{2cm}{Work process practices} \\ \cline{1-4}
Focus groups         & \textit{``Other than text, \underline{images} of different crop diseases and their symptoms would be helpful for us as many of us are \underline{not fluent in reading}. In fact, \underline{video} will be the best."}                                                                                                         & Images and videos are preferred than text because of low literacy level                    & More visuals than text         &                                                                 &                                         \\ \hline
Interviews           & \textit{``\underline{Feedback analysis} from \underline{values' lens} is really important to know the values of the female farmers. I think a \underline{new role} should be introduced for this purpose only. It can be \underline{a dedicated team or person} to analyze their feedback with both their \underline{technical and social knowledge}."} & A new role to analyze feedback from values lens with technical and non-technical knowledge & New role for feedback analysis & Establishing a dedicated team/person for values concerns        & Team structure and responsibilities     \\ \hline

\end{tabular}
}
\end{table*}
    
    \textbf{Validating the results:} We applied the member checking technique to validate the results of RQ2 as it is a popular technique to validate, verify, and assess the trustworthiness and to improve the accuracy and credibility of the findings of qualitative data analysis \cite{birt2016member, santos2017member}. As we identified 14 strategies to address Bangladeshi female farmers' values in agriculture apps (details in Subsection \ref{subsec:results_rq2}), we shared these strategies and their descriptions (2-3 sentences) with all the 13 interviewees through emails. We asked the interviewees to indicate whether they `agree' or `disagree' with each of these 14 strategies and provide reasons for their agreement or disagreement (if any). 9 of the 13 interviewees responded to us with their feedback (see the summary of their feedback \cite{replpack}). They agreed with most of the strategies. Some of them provided some constructive comments on how to improve the strategies. We made revisions after getting their feedback, but there were no major changes other than rephrasing some strategies.

\section{Results}
\label{sec:results}
This section presents the results of the analysis of the interviews with 13 app practitioners and focus groups with 20 Bangladeshi female farmers.

    \subsection{RQ1: Ratios of Bangladeshi Female Farmers' Values Reflected in Existing Agriculture Apps} 
    \label{subsec:results_rq1}
    We answered RQ1 from app practitioners' and Bangladeshi female farmers' perspectives through interviews and focus groups respectively. \autoref{table:RQ1_inw_fg} shows the results of RQ1 from the interviews as well as focus groups with corresponding illustrative quotes. It shows Bangladeshi female farmers' values that are reflected and ignored/violated in the existing agriculture apps which we named ``present" and ``missing" values respectively. We categorized a value as present value if one or more interviewees/focus groups participants think this value is reflected in the existing apps and nobody denies. Similarly, a value is categorized as missing value when one or more agree that this value is ignored/violated in the existing apps and nobody thinks this is present. \autoref{table:RQ1_No_Of_Participants} shows the number of participants from both the interviews and focus groups along with their corresponding ID who recognized a particular value. 
    
    It should be noted that through the values probing questions, we identified 15 more values (e.g. \textit{ambition}, \textit{social power}, \textit{education}, \textit{beauty}, \textit{entertainment}, \textit{family}) of Bangladeshi female farmers which the participants did not expect to be reflected in agriculture apps. As these values are not relevant with the goal of this study, we did not include them in the results. It is also worth reemphasizing that while we used Schwartz's theory of basic human values as a starting reference model to extract values, we were not restricted to values in Schwartz theory and identified other values not identified by Schwartz such as \textit{accuracy}, \textit{learning}, \textit{tidiness}, \textit{simplicity}, \textit{accessibility}, \textit{safety}, \textit{trust}, and \textit{user-friendliness}.

    
    \begin{table*}
\centering
\caption{Results of RQ1 from the interviews and focus groups: Bangladeshi female farmers' present and missing values in existing agriculture apps (INW: Interviews; FG: Focus groups; \faCheck: Present values; \faClose: Missing values; \faCommentsO: Quotes from interviews; \faComments: Quotes from focus groups)}
\label{table:RQ1_inw_fg}
\resizebox{\textwidth}{!}{%
\renewcommand{\arraystretch}{1.3}
\begin{tabular}{|P{2.3cm}|c|c|P{16.3cm}|}
\hline
\multicolumn{1}{|c|}{\textbf{Values}}                           & \multicolumn{1}{c|}{\textbf{INW}} & \multicolumn{1}{c|}{\textbf{FG}} & \multicolumn{1}{c|}{\textbf{Illustrative Quotes (INW \& FG)}}                                                                                                                                                                                                                                                                                                                                                                                                                                                                                                                                                                                                                                            \\ \hline
                                            Accuracy           & \faCheck                              &                             & {\faCommentsO{} ``The female farmers gave feedback that the information in the apps is 90\% accurate. For example, weather prediction. They said if the apps predict rain, usually it rains, otherwise at least the day remains cloudy."- \textbf{I2}}                                                                                                                                                                                                                                                                                                                                                                                                                             \\ \cline{1-4} 
                                            Capability         &                              & \faCheck                             & {\faComments{} ``The apps made us capable of solving agriculture related problems. Even, with the help of the apps, we are now able to answer all the questions regarding crop cultivation."- \textbf{G2}}                                                                                                                                                                                                                                                                                                                                                                                                                                                                                                                                                                                                                                                                                                                                                                                                                                                                                                                                                                                                                                                                                                                                                                                                                                                                                                                                                                                                                                  \\ \cline{1-4} 
                                            Honesty            & \faCheck                              &                             & {\faCommentsO{} ``In their village, there is a tendency of pesticides adulteration and sell those by slightly changing the names of some popular brands. For example, Atibrite instead of Antibrite. The app stops this dishonest activity by providing original pesticides' pictures so that the female farmers can match the names while buying."- \textbf{I12}}                                                                                                                                                                                                                                                                                                           \\ \cline{1-4} 
                                            Independence       & \faCheck                              & \faCheck                             & {\begin{tabular}[c]{@{}P{16.4cm}@{}} \faCommentsO{} ``Apps make them independent as well. For example, with the payment feature they can buy and sell anything without the help of their husbands. Through the apps, they can ask for agricultural suggestions from an expert. So, you do not need to go anywhere, you have all your information in hand."- \textbf{I11}\\ \faComments{} ``Previously we were dependent on our husbands for everything. Now, with the help of the apps, we can produce crops by ourselves, do not need to ask for money and can bear the costs of our kids' education."- \textbf{G1}\end{tabular}} \\ \cline{1-4} 
                                            Creativity         & \faCheck                              &                             & {\faCommentsO{} ``We tried to design apps in a creative way. We provided a feature of automatically crops' picture taking, like scanning, and sending to us. Another interesting feature used female farmers' own voices to record app contents and added as audio files."- \textbf{I9}}                                                                                                                                                                                                                                                                                                                                                                           \\ \cline{1-4} 
                                            Learning           & \faCheck                              & \faCheck                             & {\begin{tabular}[c]{@{}P{16.4cm}@{}} \faCommentsO{} ``I found a female farmer's Facebook status by stating ``I earned 800 BDT today by vaccinating to the cows of my neighbours". She learnt it from the app and now she is an entrepreneur."- \textbf{I3}\\ \faComments{} ``We learned about 4 ft length and 2 ft width bedding from the apps which increased our production. From apps, we also learned how to take care of livestock, when to give medicine, how to build their house etc."- \textbf{G1}\end{tabular}}                                                                                                                                                                                                       \\ \cline{1-4} 
                                            Responsibility     &                              & \faCheck                             & {\faComments{} ``With the help of the apps, now we know everything about agriculture better than our husbands. Therefore, nowadays, our husbands completely rely on us and we take all the responsibilities of cultivation."- \textbf{G1}}                                                                                                                                                                                                                                                                                                                                                                                                                                               \\ \cline{1-4} 
                                            Security           & \faCheck                              &                             & {\faCommentsO{} ``The app we developed for them was offline app, so nobody can hack. For another app, we ensured that all of them create an ID and password for the security purposes."- \textbf{I9}}                                                                                                                                                                                                                                                                                                                                                                                                                                                 \\ \cline{1-4} 
                                            Tidiness          & \faCheck                              &                             & {\faCommentsO{} ``They became very happy with the tidy way of giving information on dose of medicines and precautions to apply those medicines on crops. We provided this information with different colors and they really liked it."- \textbf{I8}}                                                                                                                                                                                                                                                                                                                                                                                                                                                                                                             \\ \cline{1-4} 
                                            Self-respect       &                              & \faCheck                             & {\faComments{} ``With the help of the apps, now we are able to assist our husbands in agriculture activities and money making. It increased our respect. Even our in-laws do not make any decision without our opinions."- \textbf{G4}}                                                                                                                                                                                                                                                                                                                                                                                                                                                                                                                   \\ \cline{1-4} 
                                            Simplicity         & \faCheck                              & \faCheck                             & {\begin{tabular}[c]{@{}P{16.4cm}@{}} \faCommentsO{} ``During the app development, we kept in mind that the app will be used by the farmers. So, the app we developed has the advantage of being easy and simple."- \textbf{I11}\\ \faComments{} ``We do not face problems with the apps as they are written in Bengali, therefore, easy to understand."- \textbf{G4}\end{tabular}}                                                                                                                                                                                                                                                                                                                                                                                                                                                                      \\ \cline{1-4} 
                                            Social recognition &                              & \faCheck                             & {\faComments{} ``I gained much knowledge from apps. About me, people usually say, ``She has much knowledge on agriculture". Sometimes they come to me to know information. I feel very proud then as I am much knowledgeable than other hundreds women."- \textbf{G1}}                                                                                                                                                                                                                                                                                                                                                                                                                                                    \\ \cline{1-4} 
                                            True-friendship    & \faCheck                              &                             & {\faCommentsO{} ``The female farmers in a village have a good friendship among themselves. We provided some recorded information by using some of their voices. The others became very excited to listen to their friends' voices in the apps."- \textbf{I8}}                                                                                                                                                                                                                                                                                                                                                                                                                                            \\ \cline{1-4} 
                                            Unity with nature  & \faCheck                              & \faCheck                             & {\begin{tabular}[c]{@{}P{16.4cm}@{}} \faCommentsO{} ``We provide weather forecast regularly through apps and also provide information regarding the measures they need to take before, during and after any natural disaster like flood, cyclone etc."- \textbf{I2}\\ \faComments{} ``From the apps, we know different disaster-resistant varieties of crops. For example, a variety of paddy called the BR1156 is flood-resistant. Now heavy rain, flood even cyclone cannot destroy our crops."- \textbf{G1}\end{tabular}}                               \\ \cline{1-4} 
 Wealth             &                              & \faCheck                             & {\faComments{} ``My husband tried fish farming for two years but he lost 60000 BDT. Then I got information from the apps on prawn farming and asked my husband to let me try this. In one year I got 120000 BDT as profit."- \textbf{G3}}                                                                                                                                                                                                                                                                                                                                                                                                                                                                                         \\ \cline{1-4}
                                            Accessibility      & \faClose                              &                             & {\faCommentsO{} ``We should work on the apps to increase the accessibility for the female farmers."- \textbf{I1}}                                                                                                                                                                                                                                                                                                                                                                                                                                                                                                                                                                                                                                                                                                                                                                                                                                                                                                                                                                                                                                                                                                                                                                                                                                                                                                                                                                                                                                                                                                                                                                                                                                                                                                 \\ \cline{1-4} 
                                            Pleasure           & \faClose                              &                             & {\faCommentsO{} ``They like illustration than text. But most of the apps are full of text; no picture, no animation, no audio-video. So, they might not find the apps interesting."- \textbf{I8}}                                                                                                                                                                                                                                                                                                                                                                                                                                                                                                                                                                      \\ \cline{1-4} 
                                            Privacy            & \faClose                              &                             & {\faCommentsO{} ``Privacy is a big concern for Bangladesh's perspective. I don't think it’s being properly emphasised yet in the apps. Users are not getting their data protection."- \textbf{I11}}                                                                                                                                                                                                                                                                                                                                                                                                                                                                                                                                \\ \cline{1-4} 
                                            Safety             &                              & \faClose                             & {\faComments{} ``In the apps, there is no information on diseases which usually attack crops in flood-prone areas. As we live in flood-prone area, we need to be aware of these diseases and their remedies so that we can take precautions before flood."- \textbf{G2}}                                                                                                                                                                                                                                                                                                                                                                                                                                                                       \\ \cline{1-4} 
                                            Tradition          & \faClose                              &                             & {\faCommentsO{} ``Developers usually follow the structures of the apps built for other developed countries. It would not work for marginalized women in Bangladesh. Their tradition is completely different and you should take care of it."- \textbf{I1}}                                                                                                                                                                                                                                                                                                                                                                            \\ \cline{1-4} 
                                            Trust              & \faClose                              &                             & {\faCommentsO{} ``Trust is a big issue for the female farmers. Why should they trust an app? At the very first place, three things should be focused while developing apps- Trust, Security and Privacy. Developers are not aware of these during apps development."- \textbf{I4}}                                                                                                                                                                                                                                                                                                                                                                                                                                                            \\ \cline{1-4}
           User-friendliness  & \faClose                              & \faClose                             & {\begin{tabular}[c]{@{}P{16.4cm}@{}} \faCommentsO{} ``None of the apps are user-friendly for Bangladeshi female farmers, they do not even care about female farmers' education level. They are full of bookish knowledge with many scientific terms. Even, developers never reach out to the farmers to get their feedback on how much user-friendly an app is."- \textbf{I2}\\ \faComments{} ``The apps would be friendly for us if you add videos on prawn diseases and the way of applying medicines on prawns."- \textbf{G3}\end{tabular}}                                                                                                                                                                                                                                                                                                 \\ \hline

\end{tabular}
}
\end{table*}
    
    \begin{table*}
\centering
\caption{No. of participants with their ID who recognized the present and missing values}
\label{table:RQ1_No_Of_Participants}
\resizebox{\textwidth}{!}{%
\renewcommand{\arraystretch}{1.3}
\begin{tabular}{|c|P{4cm}|P{5.5cm}|P{5.5cm}|}
\hline
\textbf{Present/ Missing?} & \textbf{Values} & \textbf{No. of interviewees (N= 13) recognized the value and Interviewee ID} & \textbf{No. of focus groups (N= 4) recognized the value and focus group ID} \\ \hline
\multirow{15}{*}{Present}  & Accuracy                             & 1 (I2)                                                                                                 &                                                                                                       \\ \cline{2-4} 
                           & Capability                           & -                                                                                                       & 3 (G1, G2, G4)                                                                                        \\ \cline{2-4} 
                           & Honesty                              & 1 (I12)                                                                                                & -                                                                                                      \\ \cline{2-4} 
                           & Independence                         & 4 (I6, I10, I11, I13)                                                                                  & 2 (G1, G2)                                                                                            \\ \cline{2-4} 
                           & Creativity                           & 3 (I8, I9, I10)                                                                                        & -                                                                                                      \\ \cline{2-4} 
                           & Learning                             & 6 (I1, I3, I6, I7, I8, I9)                                                                             & 3 (G1, G2, G3)                                                                                        \\ \cline{2-4} 
                           & Responsibility                       & -                                                                                                       & 1 (G1)                                                                                                \\ \cline{2-4} 
                           & Security                             & 3 (I7, I8, I9)                                                                                         & -                                                                                                      \\ \cline{2-4} 
                           & Tidiness                             & 3 (I2, I7, I8)                                                                                         & -                                                                                                      \\ \cline{2-4} 
                           & Self-respect                         & -                                                                                                       & 3 (G1, G2, G4)                                                                                        \\ \cline{2-4} 
                           & Simplicity                           & 3 (I9, I10, I11)                                                                                       & 3 (G1, G2, G4)                                                                                        \\ \cline{2-4} 
                           & Social recognition                   & -                                                                                                       & 3 (G1, G2, G4)                                                                                        \\ \cline{2-4} 
                           & True friendship                      & 3 (I8, I10, I11)                                                                                       & -                                                                                                      \\ \cline{2-4} 
                           & Unity with nature                    & 3 (I2, I7, I9)                                                                                         & 2 (G1, G2)                                                                                            \\ \cline{2-4} 
                           & Wealth                               & -                                                                                                       & 3 (G1, G2, G3)                                                                                        \\ \hline
\multirow{7}{*}{Missing}   & Accessibility                        & 3 (I1, I3, I12)                                                                                        & -                                                                                                      \\ \cline{2-4} 
                           & Pleasure                             & 2 (I1, I8)                                                                                             & -                                                                                                      \\ \cline{2-4} 
                           & Privacy                              & 3 (I8, I10, I11)                                                                                       & -                                                                                                      \\ \cline{2-4} 
                           & Safety                               & -                                                                                                       & 1 (G2)                                                                                                \\ \cline{2-4} 
                           & Tradition                            & 5 (I1, I2, I8, I9, I10)                                                                                & -                                                                                                      \\ \cline{2-4} 
                           & Trust                                & 3 (I4, I8, I11)                                                                                        & -                                                                                                      \\ \cline{2-4} 
                           & User-friendliness                    & 5 (I1, I2, I5, I7, I8)                                                                                 & 2 (G1, G3)                                                                                            \\ \hline                                       
\end{tabular}
}
\end{table*}
    
    \subsubsection{\textbf{Bangladeshi Female Farmers' Perspectives}}
    The focus groups revealed 9 present values of Bangladeshi female farmers in existing agriculture apps: \textit{capability}, \textit{independence}, \textit{learning}, \textit{responsibility}, \textit{self-respect}, \textit{simplicity}, \textit{social recognition}, \textit{unity with nature}, and \textit{wealth}. 
    For example, the participants of the focus groups G1 and G4 believed that \textit{self-respect} is present in the existing apps. As a representative statement from G4, we have: 
    
    \begin{itemize}
        \item[\faComments] \textit{``With the help of the apps, now we are able to assist our husbands in agriculture activities and money making. It increased our respect. Even our in-laws do not make any decision without our opinions."} (\textbf{G4})
    \end{itemize}
 
    On the other hand, the female farmers believed that the existing apps do not support 2 of their values: \textit{safety} and \textit{user-friendliness}. This can be exemplified by the following statement from the focus group G2 participants, indicating that \textit{safety} is missing:
    
    \begin{itemize}
        \item[\faComments] \textit{``In the apps, there is no information on diseases that usually attack crops in the flood-prone areas. As we live in flood-prone area, we need to be aware of these diseases and their remedies so that we can take precautions before flood."} (\textbf{G2})
    \end{itemize}

\begin{figure}[!htbp]
            \centering
            \includegraphics[width=0.35\textwidth]{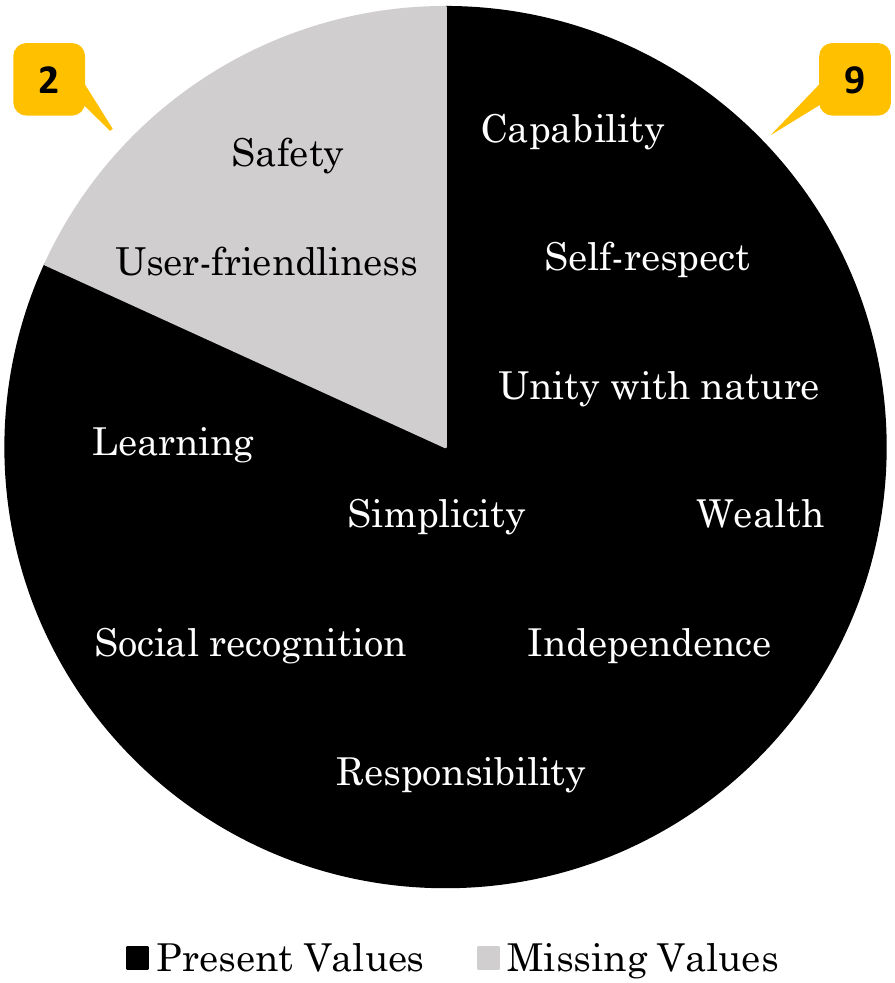}
            \caption{Ratios of Bangladeshi female farmers' present and missing values in existing agriculture apps (Results of RQ1 from the focus groups)}
            \label{fig:RQ1_focus_groups}
        \end{figure}

    It should be noted that none of the participants in other focus groups assumed that \textit{safety} is present in the apps. \autoref{fig:RQ1_focus_groups} shows the ratios of Bangladeshi female farmers' present and missing values in existing agriculture apps from their own perspectives.
    
    \subsubsection{\textbf{App Practitioners' Perspectives}}
    The interviews indicated that 10 values of Bangladeshi female farmers are reflected (present), and 6 are ignored/violated (missing) in the existing agriculture mobile apps. The present values are \textit{accuracy}, \textit{honesty}, \textit{independence}, \textit{creativity}, \textit{learning}, \textit{security}, \textit{tidiness}, \textit{simplicity}, \textit{true-friendship}, and \textit{unity with nature}. For example, one interviewee (I2) indicated that \textit{accuracy} is a value of Bangladeshi female farmers, which is present in the existing agriculture apps, and none of the interviewees thought this is missing in the apps: 
    
    \begin{itemize}
        \item[\faCommentsO] \textit{``The female farmers gave feedback that the information in the apps is 90\% accurate. For example, weather prediction. They said if the apps predict rain, usually it rains, otherwise at least the day remains cloudy."} (\textbf{I2})
    \end{itemize}
    
    On the other hand, \textit{accessibility}, \textit{pleasure}, \textit{privacy}, \textit{tradition}, \textit{trust}, and \textit{user-friendliness} were found to be missing in the existing apps. For example, none of the interviewees disagreed that \textit{privacy} is currently ignored/violated in the existing agriculture apps: 
    
    \begin{itemize}
        \item[\faCommentsO] \textit{``Privacy is a big concern for Bangladesh's perspective. I don't think it's being properly emphasised yet in the apps. Users are not getting their data protection."} (\textbf{I11})
    \end{itemize}
    
\begin{figure}[!htbp]
            \centering
            \includegraphics[width=0.35\textwidth]{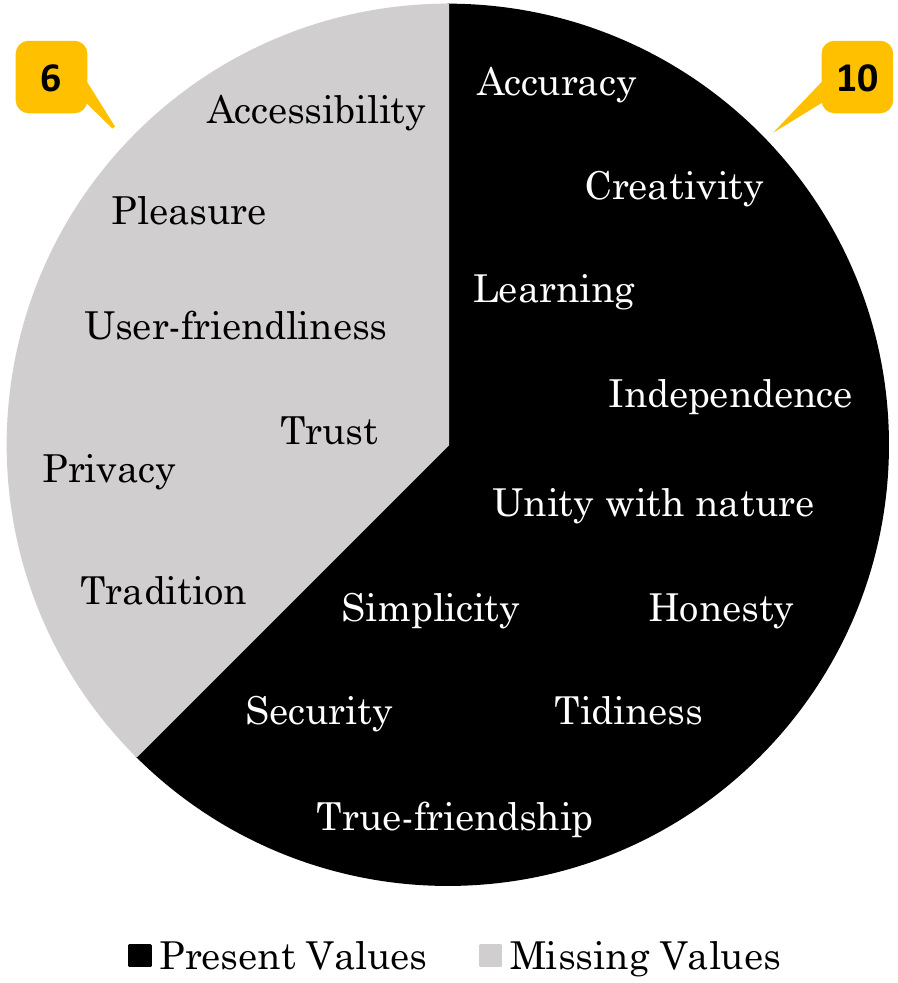}
            \caption{Ratios of Bangladeshi female farmers' present and missing values in existing agriculture apps (Results of RQ1 from the interviews)}
            \label{fig:RQ1_Interviews}
        \end{figure}

    \autoref{fig:RQ1_Interviews} shows the ratios of Bangladeshi female farmers' present and missing values in existing agriculture apps from the app practitioners' perspectives.
    
    \subsubsection{\textbf{Accumulated Results}}
    According to \autoref{table:RQ1_inw_fg}, the accumulated results from the interviews and focus groups identified 22 values of Bangladeshi female farmers which the participants expect to be reflected in the agriculture apps. Among them, 15 values are present and 7 values are missing in the existing Bangladeshi agriculture apps.
    
    \autoref{table:RQ1_inw_fg} also compares the results of RQ1 between interviews and focus groups. The participants of the interviews and focus groups indicated 6 (\textit{accuracy}, \textit{honesty}, \textit{creativity}, \textit{security}, \textit{tidiness} and \textit{true-friendship}) and 5 (\textit{capability}, \textit{responsibility}, \textit{social recognition}, \textit{wealth} and \textit{self-respect}) unique present values in existing agriculture apps respectively. \autoref{table:RQ1_inw_fg} also shows the commonalities between the results of the interviews and focus groups. The participants of the interviews and focus groups have an agreement that 4 Bangladeshi female farmers' values are present in the existing agriculture apps. They are \textit{independence}, \textit{learning}, \textit{simplicity} and \textit{unity with nature}. On the other hand, the interviewees and the focus groups' participants indicated 5 (\textit{accessibility}, \textit{pleasure}, \textit{privacy}, \textit{tradition} and \textit{trust}) and 1 (\textit{safety}) unique missing values respectively. However, there is also an agreement between the participants of the interviews and focus groups on missing values of Bangladeshi female farmers in the existing apps. Both believe that \textit{user-friendliness} is missing in the existing apps.
        
    \subsection{RQ2: Strategies to Address Bangladeshi Female Farmers' Values in Agriculture Apps}
    \label{subsec:results_rq2}
\begin{figure*}[!htbp]
                \centering
                \includegraphics[width=1\textwidth]{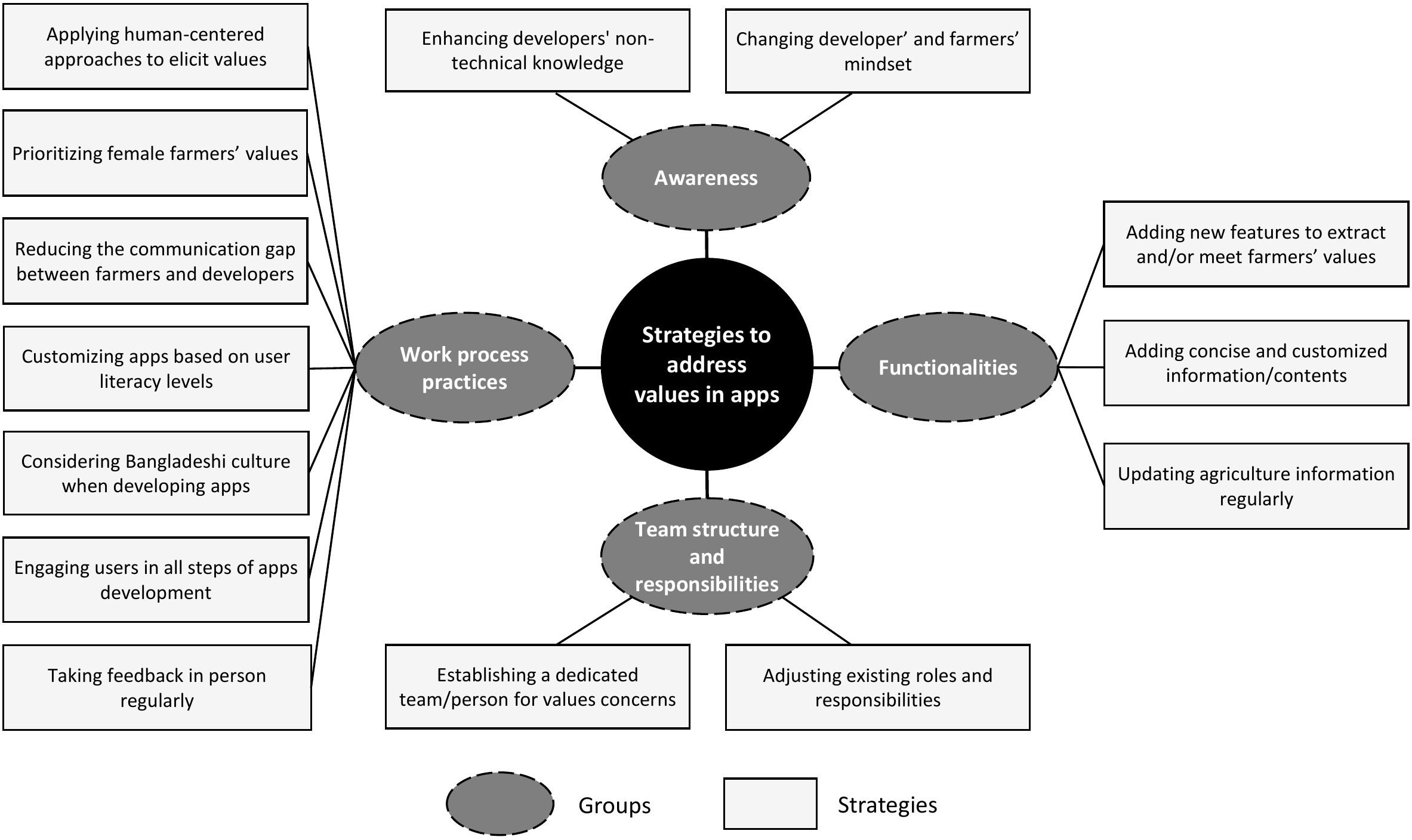}
                \caption{Results of RQ2 from the interviews and focus groups: The strategies to address Bangladeshi female farmers' values in agriculture apps}
                \label{fig:RQ2_INW_FG}
            \end{figure*}

    As with RQ1, we answer RQ2 from the perspectives of app practitioners and Bangladeshi female farmers with the aim to explore the possible strategies to address Bangladeshi female farmers' values in agriculture mobile apps. \autoref{fig:RQ2_INW_FG} shows the results of RQ2. The analysis of the interviews and focus groups data led to 14 strategies to address the values of Bangladeshi female farmers in agriculture mobile apps. These 14 strategies can be classified into four groups: `functionalities', `awareness', `work process practices', and `team structure and responsibilities'. Details of each group, its strategies, and representative quotations are given in the following sub-sections. We use three icons to show if a given strategy has emerged from the interviews (\faSlideshare{}) or focus groups (\faGroup{}) or both the interviews and focus groups (\faGroup{} \faSlideshare{}).
            
    \subsubsection{\textbf{Functionalities}} 
    We define `functionalities' as ``new features/functional units that need to be developed, and information that needs to be added and updated in the existing apps". The interviewees and focus groups' participants believed more functionalities need to be considered to the existing agriculture apps to address Bangladeshi female farmers' values. Specifically, they recommended new features to extract and/or meet farmers' values. Furthermore, they also recommended adding concise and customized information and content in the apps and regularly updating the information.
    
    \faGroup{} \faSlideshare{} \textbf{Strategy 1: Adding New Features to Extract and/or Meet Farmers' Values.}
    We found references in the interviews and the focus groups, indicating that new features need to be added to the apps to extract farmers' values. For example, they recommended new features to enable female farmers to express their feedback and needs through which their values can be extracted. From the discussions that happened in G1, we found that it is difficult for the farmers to express their needs and expectations from the apps before they start using the apps. Hence, it is challenging for the developers to realize the female farmers' actual needs from the apps as well. While the existing apps do not provide any opportunity for the farmers to express their necessities, it was found that it would be helpful if apps add features enabling the female farmers to provide their needs and feedback when using the apps.
    
    \begin{itemize}
        \item[\faCommentsO] \textit{``It is important to add a feature on providing feedback so that the users can immediately inform what problems they are facing and what are their expectations. It would inspire more female farmers to use the app and also help developers understand the users' values."} (\textbf{I5})
    \end{itemize}
    
     The participants of the interviews and focus groups also recommended several examples of required app features that meet Bangladeshi female farmers' values. For example, adding features for online buy/sell, market price analysis, and profit assessment, adding a hotline number for consultancy, providing weekly voice messages on how to take care of crops, adding videos on crop diseases, prevention, and remedies for diseases. 

    \begin{itemize}
        \item[\faCommentsO] \textit{``If a feature can be provided in the existing apps to call the consultants, the users could ask all their queries directly by calling that hotline number."} (\textbf{I11})
        \item[\faComments] \textit{``Although the apps we use provide much information on agriculture, but we do not find any information regarding crop diseases and their solutions. If these can be provided through videos, we would be able to match our crops with the videos and take steps accordingly."} (\textbf{G4})
    \end{itemize}
    
    \faGroup{} \faSlideshare{} \textbf{Strategy 2: Adding Concise and Customized Information/Contents.}
    The participants of the interviews and focus groups also suggested adding concise and customized information/contents in the apps to address farmers' values. They also shared some examples such as information on post-harvesting and processing, weather prediction, agriculture advice corresponding to weather information, applying fertilizers, crop preservation, varieties of seeds, soil fertility, and pest control. For example, regarding the weather information and crop preservation, the participants said:    
    \begin{itemize}
        \item[\faCommentsO] \textit{``Besides the regular weather phenomena, sometimes Bangladesh faces extreme natural disaster such as, flood, cyclone, drought and so on. Therefore, it is necessary to provide information on weather prediction and agriculture advise to keep the productions safe during any unwanted natural disaster."} (\textbf{I2})
        \item[\faComments] \textit{``If we want to sell our crops after 6 months for getting high price, we usually try to dry it properly before preservation but still it gets wet. Therefore, it would be helpful for us if we get information on how to preserve crops."} (\textbf{G1})
    \end{itemize}
    
    The interviewees and focus groups' participants encouraged not to add information on different crops in an app, rather app with the information of a single crop would be simple and easy to use. For example, it was mentioned by the participants of G4 that they had faced problems in understanding the apps that have too much information on different crops. They found the app only for prawn farming more useful than the apps with information on many crops.
    
    \begin{itemize}
        \item[\faCommentsO] \textit{``Too much information usually makes an app very messy, it can also confuse the users. I never support adding information of different crops in a single agriculture app. Rather we should customize apps based on different crops which would be more helpful for the female farmers."} (\textbf{I7})
    \end{itemize}
    
    \faGroup{} \faSlideshare{} \textbf{Strategy 3: Updating Agriculture Information Regularly.}
    We got this recommendation from 9 interviewees (I1, I2, I3, I5, I6, I7, I9, I11, and I13) and two focus groups (G1 and G2). They believed that regular updates of information based on Bangladeshi female farmers' requirements and values are necessary to make the apps more helpful for them. 
    
    \begin{itemize}
        \item[\faCommentsO] \textit{``Developing apps should not be our only task, we should have some updating mechanisms for the apps. For example, a pesticide is working well now but after 6 months Government might ban it for some reasons. If we do not update this information on the apps, the farmers will continue using it."} (\textbf{I3})
    \end{itemize}
    
    \begin{itemize}
        \item[\faComments] \textit{``Crop diseases change with time and weather which we do not know how to deal with. Therefore, updated information needs to be added regularly in the apps."} (\textbf{G1})
    \end{itemize}
    
    \subsubsection{\textbf{Awareness}}
    `Awareness' is referred to as ``concern about values in apps and gaining knowledge accordingly". Only the interviewees proposed the recommendations for `awareness', and no recommendations emerged from the focus groups. The interviewees provided the two following recommendations.
    
    \faSlideshare{} \textbf{Strategy 4: Enhancing Developers' Non-technical Knowledge.}
    In most cases, the development team is not familiar with the term, ``values". Hence, it is challenging for them to realize the importance of addressing values in apps. Therefore, it was recommended that developers should be trained to enhance their knowledge of values concepts and the necessity of addressing values in apps.
    
    \begin{itemize}
        \item[\faCommentsO] \textit{``Bookish knowledge can enhance the technical knowledge, but to understand the word ``values", developers should be trained on values concepts and the importance of addressing values in apps."} (\textbf{I4})
    \end{itemize}
    
     It is also difficult to identify the values of farmers and explore the process of addressing their values in agriculture apps with technical knowledge only. Therefore, developers also need to enhance their non-technical knowledge, particularly psychological and social knowledge. 
    
    \begin{itemize}
        \item[\faCommentsO] \textit{``Developers' should build social knowledge as well because they always start thinking from technical perspectives where thinking from social perspectives is also important to address female farmers' values in apps."} (\textbf{I4})
    \end{itemize}
    
    \faSlideshare{} \textbf{Strategy 5: Changing Developers' and Farmers' Mindset.}
    It is a usual practice that developers are not involved directly in farmers' values extraction and/or addressing those values in app development. The interviewees suggested developers think from farmers' perspectives during app development. The interviewees believed that this is only possible if the developers build a mindset to help farmers.
    
    \begin{itemize}
        \item[\faCommentsO] \textit{``We never think from out of the box. We just implement what we have learnt/studied but addressing values in apps would not work in this way. We should build our mindset of helping female farmers."} (\textbf{I1})
    \end{itemize}
    
    The interviewees also had suggestions for the farmers to address their values in apps. 
    Though many of the female farmers have been trained to use smartphones, particularly agriculture mobile apps, they were still very hesitant to use apps. According to the interviewees, the farmers should gradually develop a mindset of using agriculture apps.
    
    \begin{itemize}
        \item[\faCommentsO] \textit{``Using apps is not something we can force the female farmers to do. It is all about mindset of using apps. They need to realize that if they spend 20 BDT on internet to use the apps, they can save thousands BDT from their production. This mindset is not going to be developed overnight, we can expect changing it gradually."} (\textbf{I1})
    \end{itemize}

    \subsubsection{\textbf{Work Process Practices}} 
    We define 'work process practices' as ``the practices that can be adopted throughout the development life-cycle to enable or support inclusion of human values in apps''.  
    We found two recommendations from both the participants of the interviews and focus groups. The interviewees also indicated another five recommendations under this theme.
    
    \faSlideshare{} \textbf{Strategy 6: Applying Human-Centered Approaches to Elicit Values.}
    The interviewees proposed several human-centered approaches to elicit values, such as observations, asking non-scientific questions, employing a bottom-up approach, employing the design thinking method, and using A/B testing. Notably, all of the interviewees recommended adopting a bottom-up approach to elicit values.

    \begin{itemize}
        \item[\faCommentsO] \textit{``Usually the developers think user-centered design is the holy grail to address values in apps. But I recommend an updated version of user-centered design which is design thinking. It can be applied to understand the values of the users, to realize the challenges involved and to design the prototypes of apps accordingly."} (\textbf{I13})
    \end{itemize}
    
    \begin{itemize}
        \item[\faCommentsO] \textit{``I think we can use A/B testing to identify farmers' values. For example, farmers use call center to discuss their agriculture issues with the consultants. As the way of talking reflects people's values, we can elicit farmers' values by considering two variants of A/B test: Way of talking and values".} (\textbf{I13})

    \end{itemize}
    
    \faSlideshare{} \textbf{Strategy 7: Prioritizing Female Farmers' Values.}
The interviewees recommended prioritizing values based on the requirements of the farmers. For example, if a farmer has a list of 20 desired values, a few are the most important to them. During app development, addressing the most important values should be given priority. For instance, I4 thinks \textit{security}, \textit{privacy}, and \textit{trust} are the most demanding values for the farmers. Therefore,  these three values should be addressed first in apps and then the values of lesser priority.

    \begin{itemize}
        \item[\faCommentsO] \textit{``If you find many values of the female farmers, there must be some values which are the most important to them. Therefore, value prioritization is necessary before developing apps."} (\textbf{I13})
    \end{itemize}
    
    \faSlideshare{} \textbf{Strategy 8: Reducing The Communication Gap Between Farmers and Developers.}
    It is not common for developers to undertake `field visits', which certainly results 
    in a communication gap between the developers and the farmers. As a result, the developers do not realize the values of the farmers and, therefore, cannot address their values in apps as well. Four interviewees (I1, I3, I4, and I5) believed that reducing the communication gap between female farmers and developers could help understand farmers' values and address those in apps. 
    
    \begin{itemize}
        \item[\faCommentsO] \textit{``There is a huge gap between the community people and the technical people. As the developers do not visit the female farmers, their requirements reach to the developers through different people. Therefore, there are different information layers between the developers and the community people and the actual requirements are modified because of these multiple information layers."} (\textbf{I5})
    \end{itemize}
    
    \faGroup{} \faSlideshare{} \textbf{Strategy 9: Customizing Apps Based on User Literacy Levels.} As most of the Bangladeshi female farmers have relatively low levels of formal education, the apps should be designed based on the farmers' literacy levels and customized in a way that is easier for them to use. According to I11, this should be the highest priority when developing apps for them. The apps should be designed so that 90\% of female farmers can use them without any trouble.
    
    \begin{itemize}
        \item[\faCommentsO] \textit{``Maybe too much information, complicated design and too much colors can make the apps attractive but not usable. My recommendation is to design the apps in a way that it feels simple to use."} (\textbf{I8})
    \end{itemize}
    
Two interviewees (I11 and I13) and three focus groups (G2, G3, and G4) recommended that visual imagery, rather than text, should be used by the apps  because of  the low literacy level of the farmers.
    
    \begin{itemize}
        \item[\faComments] \textit{``Other than text, images of different crop diseases and their symptoms would be helpful for us as many of us are not fluent in reading. In fact, video will be the best."} (\textbf{G3})
    \end{itemize}
    
    \faSlideshare{} \textbf{Strategy 10: Considering Bangladeshi Culture When Developing Apps.}
    Respecting local cultural norms is essential for the people of all cultural settings. 
    The apps that are not aligned with Bangladeshi culture may not be considered satisfactory by Bangladeshi female farmers.
    
    \begin{itemize}
        \item[\faCommentsO] \textit{``If you want to make an app for the female farmers in Bangladesh, you cannot follow the structure of an app developed for other countries even for other societies. You need to respect their culture while developing apps for them. The UI/UX should be different, the interaction should be different, even the language and contents should be different."} (\textbf{I1})
    \end{itemize}
    
    \faSlideshare{} \textbf{Strategy 11: Engaging Users in All Steps of Apps Development.}
    The interviewees recommended the development teams to include users in all steps of the software development life cycle, which will help them understand the users' values and incorporate those values in app development.

    \begin{itemize}
        \item[\faCommentsO] \textit{``We should include the users from the very beginning of app development. In fact, I think we should include them in all steps of software development life cycle to develop apps based on their values."} (\textbf{I4})
    \end{itemize}
    
    \faGroup{} \faSlideshare{} \textbf{Strategy 12: Taking Feedback in Person Regularly.}
    This recommendation came from the participants of both interviews and focus groups. Besides adding features for collecting feedback, they also felt the necessity of collecting feedback in person from the users. 10 interviewees except for I7, I8, and I12 recommended taking feedback regularly in person to improve the apps based on Bangladeshi female farmers' values.

    \begin{itemize}
        \item[\faCommentsO] \textit{``We always think that we developed an app and our task is over. This is not the right way. We need to visit and talk to the users in every 2-3 months to know their feedback and should update the app accordingly."} (\textbf{I9})
    \end{itemize}
    
      Only the participants of G1 deemed regular feedback taking in person would be much helpful. According to them, if developers or field facilitators come to them and discuss what they need, that would help the developers understand what actually their requirements are and work accordingly.

    \subsubsection{\textbf{Team Structure and Responsibilities}} The last group is `team structure and responsibilities'. The interviewees recommended adding new roles, and both the participants of the interviews and focus groups recommended adjusting the structure and responsibilities of the existing teams to address female farmers' values in apps.
    
    \faSlideshare{} \textbf{Strategy 13: Establishing a Dedicated Team/Person for Values Concerns.}
    This recommendation came from the interviewees, suggesting a new role or team should take regular feedback, analyze the feedback from the values' lens, and work on those accordingly while developing apps.
    
    \begin{itemize}
        \item[\faCommentsO] \textit{``Feedback analysis from values' lens is really important to know the values of the female farmers. I think a new role should be introduced for this purpose only. It can be a dedicated team or person to analyze their feedback with both their technical and social knowledge."} (\textbf{I4})
    \end{itemize}
    
    \faGroup{} \faSlideshare{} \textbf{Strategy 14: Adjusting Existing Roles and Responsibilities.}
    The interviewees thought that the existing app development team members could take more responsibilities to address female farmers' values in agriculture apps. For example, the field facilitators usually talk to the female farmers and convey their requirements, feedback, and all the necessary information to the app development team. However, the interviewees believed that field visits by product managers, developers, and other stakeholders would better understand farmers' values and address those in apps. Furthermore, the product owner and UX designer need to understand the term "values" and address those values in apps.  The interviewees named such new or adjusted responsibilities "value responsibilities".
    
    \begin{itemize}
        \item[\faCommentsO] \textit{``I don't think any new role need to be introduced to design apps that respect human values. I believe the existing roles such as product owner and UX designer should be able to take more responsibilities to address values in apps."} (\textbf{I13})
    \end{itemize}
    The focus groups' participants did not suggest any new role to address values in apps. Instead, G1, G2, and G3 suggested new responsibilities for the development team members. They requested team members (e.g., developers) to visit the field and talk to them to understand their values and develop apps accordingly.
    
    \begin{itemize}
        \item[\faComments] \textit{``If the people who develop apps come to us, we can raise our concerns directly to them."} (\textbf{G1})
    \end{itemize}
    
\section{Discussion and Implications}
\label{sec:discussion_v2}
This section discusses the findings of this study as well as the implications for software engineering research and practice.

    \subsection{Prioritizing Bangladeshi Female Farmers' Values}
    \label{subsec:PrioritizingValues}

    This section compares the findings of RQ1 with previous related work. Out of 22 values of Bangladeshi female farmers which our participants expect to be reflected in agriculture apps, the comparison indicates that 13 values are reported in \cite{schwartz1992universals}, \cite{shams2021measuring}, and \cite{shams2020society}. They are \textit{capability}, \textit{honesty}, \textit{responsibility}, \textit{true-friendship}, \textit{independence}, \textit{creativity}, \textit{self-respect}, \textit{privacy}, \textit{security}, \textit{unity with nature}, \textit{tradition}, \textit{pleasure}, and \textit{wealth}.
    
    From the 22 values (15 present values and 7 missing values) of Bangladeshi female farmers identified from this study (see \autoref{table:RQ1_inw_fg}), 14 values can be found in Schwartz's theory of basic human values \cite{schwartz1992universals}. They are \textit{capability}, \textit{honesty}, \textit{responsibility}, \textit{true-friendship}, \textit{independence}, \textit{creativity}, \textit{self-respect}, \textit{privacy}, \textit{security}, \textit{unity with nature}, \textit{tradition}, \textit{social recognition}, \textit{wealth}, and \textit{pleasure}. According to Schwartz, these values can be placed under 8 main value categories: Achievements, Benevolence, Self-direction, Security, Universalism, Tradition, Power, and Hedonism. However, among 8 of these main value categories, 6 (Achievements, Benevolence, Self-direction, Security, Universalism, and Tradition) are the most preferred values of Bangladeshi female farmers according to Shams et al. \cite{shams2021measuring}. At the same time, Shams et al. observed that Power and Hedonism are the least important values for Bangladeshi female farmers \cite{shams2021measuring}.
    
    Further to this, 10 out of 22 values are similar to the values of end-users of Bangladeshi agriculture mobile applications identified from app reviews  \cite{shams2020society}. They are \textit{capability}, \textit{honesty}, \textit{independence}, \textit{creativity}, \textit{responsibility}, \textit{security}, \textit{wealth}, \textit{pleasure}, \textit{privacy}, and \textit{tradition}.

    Therefore, the comparison of our study with previous related work indicates that 13 values are common for Bangladeshi female farmers when using Agriculture apps. 
    Hence, while we recommend app developers address all the 22 values during agriculture apps development for Bangladeshi female farmers, it is essential to prioritize these 13 common values to ensure that none of these 13 values are left out during development. This is in line with one of the strategies we identified to address Bangladeshi female farmers' values in agriculture mobile apps, \textit{``Strategy 7: prioritizing female farmers' values"}.
    
    \begin{tcolorbox}[colback=gray!5!white,colframe=gray!75!black,title=Implication 1]
        \justify 
        App developers need to address and prioritize the following values when developing agriculture apps for Bangladeshi female farmers: \textit{capability}, \textit{honesty}, \textit{responsibility}, \textit{true-friendship}, \textit{independence}, \textit{creativity}, \textit{self-respect}, \textit{privacy}, \textit{security}, \textit{unity with nature}, \textit{tradition}, \textit{pleasure}, and \textit{wealth}. 
    \end{tcolorbox}
    

    \subsection{Highly Emphasized Strategies to Embed Values in Apps Development}
    \label{subsec:PrioritizingStrategies}
    
    This section compares the findings of RQ2 with previous related work. Out of the 14 strategies to address Bangladeshi female farmers' values in agriculture apps, we argue that the following 5 strategies are close to some of the strategies reported in \cite{hussain2020human, hussain2021can}: \textit{``Strategy 1: Adding new features to extract and/or meet farmers' values"}, \textit{``Strategy 6: applying human-centered approaches to elicit values"}, \textit{``Strategy 12: taking feedback in person regularly"}, \textit{``Strategy 13: establishing a dedicated team/person for value concerns"}, and \textit{``Strategy 14: adjusting existing roles and responsibilities"}.
    
    The strategy proposed by Hussain et al. \cite{hussain2020human}, ``develop a sense of empathy with users", is close to our strategy, \textit{```Strategy 1: Adding new features to extract and/or meet farmers' values"}. Similarly,  the strategies ``apply user-centered elicitation techniques to implement human values", and ``A/B tests to validate assumptions about user values" \cite{hussain2020human} can be likened to our strategy, \textit{``Strategy 6: applying human-centered approaches to elicit values"}. Finally, ``prototyping iteratively and using customers' feedback", ``use attentional experience for user feedback on values implementation", and ``qualitative feedback on emotions of system use" \cite{hussain2020human} are close to our strategy, \textit{``Strategy 12: taking feedback in person regularly"}.
    
    Similarly, we also found a few similarities of our findings with another study by Hussain et al. \cite{hussain2021can}. The strategy, ``seek user feedback beyond usability, functionality to capture user experiences of their values expectations" is quite similar to our strategy, \textit{``Strategy 12: taking feedback in person regularly"}. Hussain et al. \cite{hussain2021can} proposed some new roles for value concerns such as values champions, values mentors, values translators, values promoters, and values officers, which are similar to our strategy, \textit{``Strategy 13: establishing a dedicated team/person for value concerns"}. Similarly, they suggested adding some responsibilities to existing roles such as product owners, business owners, and release train engineer, which is close to our another strategy, \textit{``Strategy 14: adjusting existing roles and responsibilities"}.
    
    While we recommend app developers adopt all the 14 strategies we identified to include Bangladeshi female farmers' values in agriculture apps developments, we argue that 5 of them are of greater importance because they have also been suggested by previous related work. Therefore, the app developers need to ensure that none of these 5 strategies are left out during agriculture apps development for Bangladeshi female farmers.
    
    \begin{tcolorbox}[colback=gray!5!white,colframe=gray!75!black,title=Implication 2]
        \justify 
        App developers need to emphasize the following strategies when developing agriculture apps for Bangladeshi female farmers: \textit{``Adding new features to extract and/or meet farmers' values"}, \textit{``applying human-centered approaches to elicit values"}, \textit{``taking feedback in person regularly"}, \textit{``establishing a dedicated team/person for value concerns"}, and \textit{``adjusting existing roles and responsibilities"}.
    \end{tcolorbox}
\section{Threats to Validity}
\label{sec:ttv}

The trustworthiness of research depends on the credibility, confirmability, dependability, and transferability of that research \cite{cruzes2011recommended}. This section discusses the possible threats arising from this research according to these four validation criteria.

    \subsection{Credibility}
    The potential threats to the credibility of this research could arise from the data collection approach, designing the questionnaires for interviews and focus groups, participants' selection approach, and the sample size of the focus groups.
    
   We believe the data collection process from different sources (data source triangulation \cite{triangulation2014use}) increased the plausibility of our findings. Particularly, data triangulation using focus groups and interviews in qualitative investigation develops a comprehensive understanding of the phenomenon of interest \cite{triangulation2014use}.
    
    To mitigate the threats raised from the questionnaires for interviews and focus groups, we used open-ended questions. We asked several follow-up questions according to the responses of the participants. We also allowed the participants to share their experiences and any stories related to agriculture apps.
    
    To limit the impacts of the threats caused by the selection of participants, we shared our participants' selection criteria and the objective of this study with the senior employees of Oxfam Bangladesh. They helped us recruit the right participants. For interviews, we also conducted Snowball Sampling Method to mitigate the possible threats.
    
    Another potential threat to the credibility of this research is the sample size of the focus groups. We conducted 4 focus groups with 20 participants in this study, whereas Millward suggested 6 to 12 focus groups with a minimum of 21 participants \cite{millward1995focus} to reach saturation. As the number of participants and number of focus groups of this study is quite close to the suggestions from the literature, we believe it reduced the possible threats to credibility. Furthermore, we also collected data from 13 interviews. Research shows that 12 interviews are sufficient to reach the saturation for qualitative research \cite{sim2018can}. Although researchers believe that sample size is an important issue to get credible results in both qualitative and quantitative research, some researchers believe that there are other essential elements in research as well other than the sample size only \cite{sim2018can}. According to Emmel, ``it is not the number of cases that matters, it is what you do with them that counts" \cite{emmel2013sampling}.
    

    \subsection{Confirmability}
    \label{ttv_confirmability}
    The potential threats to the confirmability of this study could arise from the absence of initial investigator triangulation, recognition of a present or missing value by one participant only, the absence of validation of the findings of RQ1, and the concerns raised from the member checking process.
    
    As the initial coding process of this research was accomplished by one analyst (the first author) only, we accept that it might cause a significant threat to the confirmability of this study. However, the first author tried to mitigate this threat by repeating the data analysis thrice to confirm the intra-rater reliability \cite{ergai2016assessment}. Moreover, after the initial investigation, several rounds of discussions happened among all the authors to develop the final codebook. In addition, we also conducted member checking to mitigate the threat and validate the results (findings of RQ2).
    
    We also accept that another potential threat to the confirmability of this study could arise from the findings of RQ1. We identified two values (\textit{accuracy} and \textit{honesty}) mentioned by one interviewee and another two values (\textit{responsibility} and \textit{safety}) by one focus group only. To mitigate this threat and the threats introduced from the absence of validation of the findings of RQ1, we compared our results with the previous related work (Subsection \ref{subsec:PrioritizingValues}), and we found that our findings are close to the related work.

    From the member checking process to validate the strategies of addressing values in apps, we got a few concerns that could be potential threats to confirmability. For example, three participants (I2, I6, and I12) raised concerns about feasibility from financial perspectives to implement some strategies (Strategy 1, Strategy 11, Strategy 12, Strategy 13). We agree with their concerns, and in the future, we will work on reducing the associated costs to implement these strategies.
    

    \subsection{Dependability}
    The ill-defined nature of human values could introduce the possible threat to the dependability of this research \cite{perera2020study}. Human values are an abstract topic, and there are no definitions of human values from mobile apps/software engineering perspectives. 
  
    It can be argued that there is a possible threat to interpret values differently by the participants during interviews and focus groups. However, in focus groups, we never used the word ``values" to ask the participants, limiting the risks of possible threats. For interviews, we mitigated this threat by explaining the term ``values" with proper examples and described Schwartz's theory of basic human values before the interview session started.
    
    It can also be argued that the researchers might interpret values differently during data analysis according to their understanding of values. However, we believe the risk is low as all the authors of this research are highly experienced in human values in software engineering.
    

    \subsection{Transferability}
    As the end-users of this research are Bangladeshi female farmers, there is a possible threat to the transferability of this study. As different groups of end-users might have different values when using different apps, it can be argued that our results cannot be generalized for all the end-users of mobile apps. However, we believe our results can be used for the users and apps of other developing countries like Bangladesh. Furthermore, after the comparison of the results of RQ1 and RQ2 with the previous related work in Subsection \ref{subsec:PrioritizingValues} and \ref{subsec:PrioritizingStrategies}, we found a few similarities of our findings with the other related work. Therefore, we believe that our methodology can be replicated for the users and apps in different cultural settings.
\section{Conclusions and Future Work}
\label{sec:conclusion}
We conducted a mixed-methods empirical study with the goal to identify the extent to which existing mobile apps reflect the values of their end users and to explore the possible strategies to address their values in apps. For this purpose, we conducted interviews with 13 app practitioners and focus groups with 20 Bangladeshi female farmers who use agriculture apps in their daily agricultural activities. The accumulated results identified 22 values of Bangladeshi female farmers which the participants expect to be reflected in the agriculture apps. Among them, 15 values are already reflected (present) and 7 values are ignored/violated (missing) in the existing Bangladeshi agriculture apps. Particularly, from interviews, we identified 10 present (e.g. \textit{accuracy}, \textit{creativity}, \textit{independence} etc.) and 6 missing values (e.g. \textit{accessibility}, \textit{pleasure}, \textit{trust} etc.) of Bangladeshi female farmers in the existing agriculture apps whereas the focus groups indicated 9 present (e.g. \textit{capability}, \textit{simplicity}, \textit{self-respect} etc.) and 2 missing values (e.g. \textit{safety} and \textit{user-friendliness}). We also explored 4 groups (`functionalities', `awareness', `work process practices', and `team structure and responsibilities') with 14 strategies (e.g. \textit{``adding concise and customized information/contents"}, \textit{``enhancing developers' non-technical knowledge"}, \textit{``applying human-centered approaches to elicit values"}, \textit{``adjusting existing roles and responsibilities"} etc.) from both the interviews and focus groups to address Bangladeshi female farmers' values in agriculture apps. 

We believe these findings can be used as guidelines for app developers to develop apps that respect the values of their end-users. This research is also an example to conduct more research to address human values in software.

Our future work will be focused on the collaboration with app developers to implement the findings of this research during app development. During implementation, we will also work on how to reduce possible costs to implement the proposed strategies. Furthermore, we also aim to increase the sample size of this research by including male farmers and other end-users of apps from different cultural settings to ensure the transferability of this research.
\bibliographystyle{IEEEtran}
\bibliography{References}

\end{document}